\documentclass[12pt]{article}
\usepackage{amssymb}
\usepackage{epsf}
\usepackage{epsfig}
\usepackage{cite}
\newcommand{\lsim}{
\mathrel{\hbox{\rlap{\hbox{\lower4pt\hbox{$\sim$}}}\hbox{$<$}}}}

\newcommand{\gsim}{
\mathrel{\hbox{\rlap{\hbox{\lower4pt\hbox{$\sim$}}}\hbox{$>$}}}}

\begin{document}
\begin{titlepage}
\vspace*{-0.5truecm}

\begin{flushright}
CERN-PH-TH/2007-092\\
\end{flushright}

\vspace*{1.3truecm}

\begin{center}
\boldmath
{\Large{\bf 
Exploring CP Violation and Penguin Effects through $B^0_d\to D^+D^-$ and 
$B^0_s\to D^+_sD^-_s$
}}
\unboldmath
\end{center}

\vspace{0.9truecm}

\begin{center}
{\bf Robert Fleischer}

\vspace{0.5truecm}

{\sl Theory Division, Department of Physics, CERN, CH-1211 Geneva 23,
Switzerland}

\end{center}

\vspace{0.6cm}
\begin{abstract}
\vspace{0.2cm}\noindent
The decay $B^0_d\to D^+D^-$ offers an interesting probe of CP violation, but requires 
control of penguin effects, which can be done through $B^0_s\to D^+_sD^-_s$
by means of the $U$-spin flavour symmetry of strong interactions. Recently, the 
Belle collaboration reported indications of large CP violation in the 
$B^0_d$ decay, which were, however, not confirmed by BaBar, and first signals 
of the $B^0_s$ channel were observed at the Tevatron. In view of these developments
and the quickly approaching start of the LHC, we explore the allowed region 
in observable space for CP violation in $B^0_d\to D^+D^-$, perform theoretical 
estimates of the relevant hadronic penguin parameters and observables, and address 
questions both about the most promising strategies for the extraction of CP-violating 
phases and about the interplay with other measurements of CP violation and the search 
for new physics. As far as the latter aspect is concerned, we point out that the 
$B^0_{q} \to D_{q}^+D^-_{q}$ system provides a setting for the determination of the 
$B^0_q$--$\bar B^0_q$ mixing phases ($q\in\{d,s\}$) that is complementary to the 
conventional $B^0_d\to J/\psi K_{\rm S}$ and $B^0_s\to J/\psi \phi$ modes with 
respect to possible new-physics effects in the electroweak penguin sector.
\end{abstract}

\vspace*{0.5truecm}
\vfill
\noindent
May 2007

\end{titlepage}

\thispagestyle{empty}
\vbox{}
\newpage

\setcounter{page}{1}

\begin{figure}
\centerline{
 \includegraphics[width=5.8truecm]{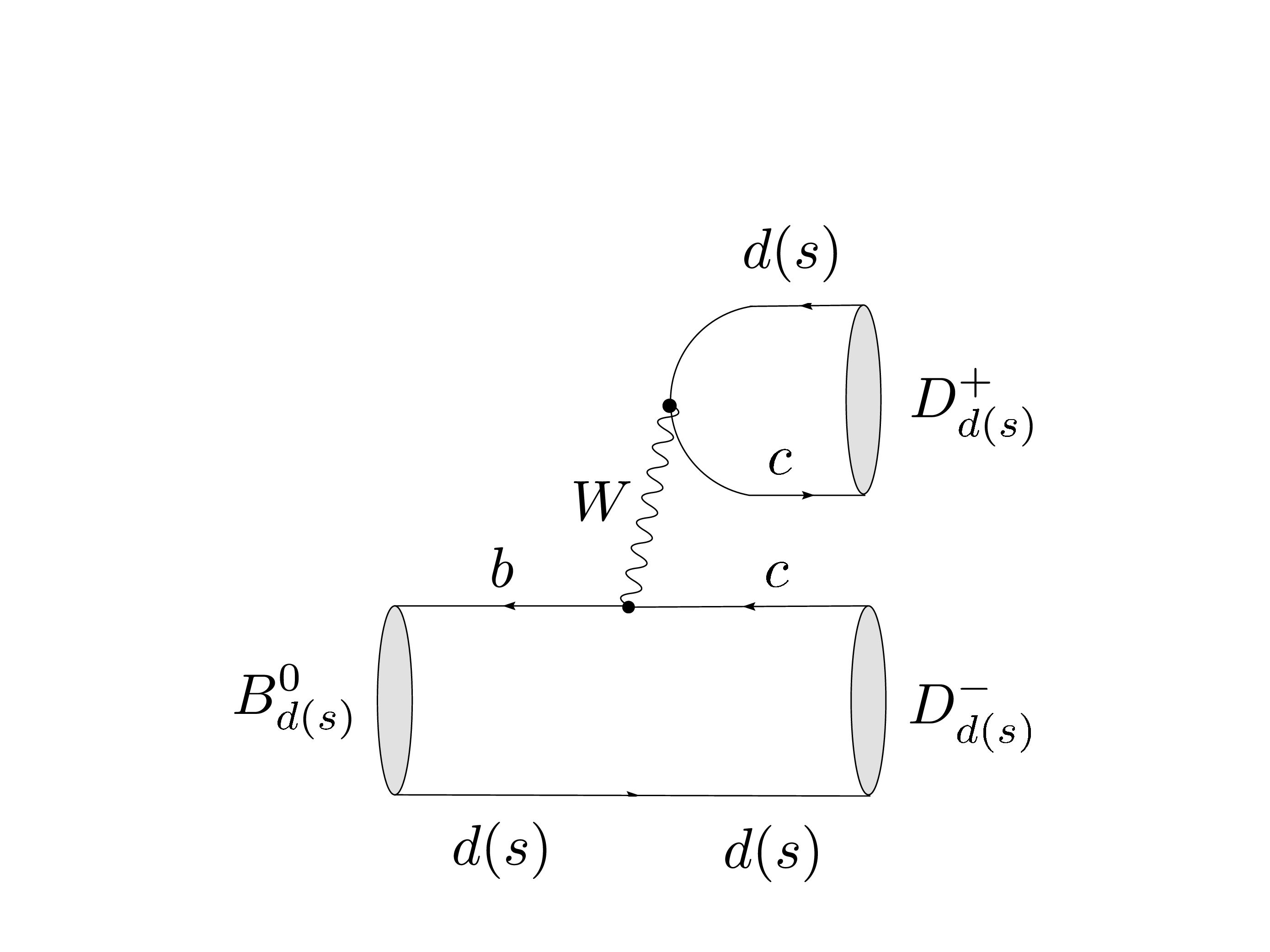}
 \hspace*{0.5truecm}
 \includegraphics[width=6.4truecm]{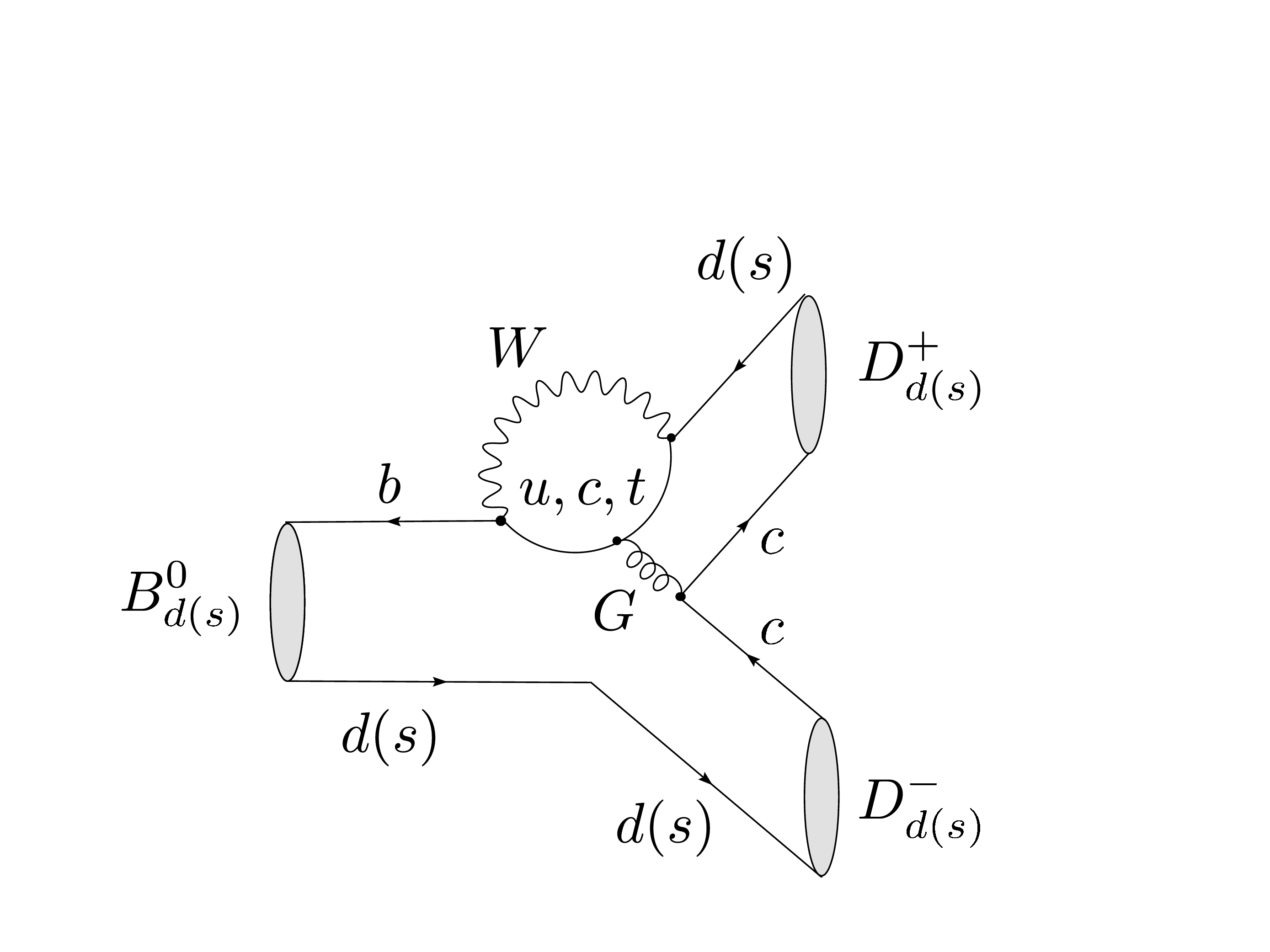}  
 }
 \vspace*{-0.1truecm}
\caption{Tree and penguin topologies contributing to the $U$-spin-related 
$B^0_d\to D^+D^-$ and $B^0_s\to D_s^+D_s^-$  decays.}\label{fig:BDD-diag}
\end{figure}

\section{Introduction}\label{sec:intro}
Decays of $B$ mesons are subject of intensive investigations. Thanks to the 
interplay between a lot of theoretical work and the data from the $e^+e^-$ 
$B$ factories with their detectors BaBar (SLAC) and Belle (KEK)
as well as from the Tevatron (FNAL), valuable insights into CP violation 
could be obtained during the recent years. In the Standard Model (SM), 
this phenomenon is closely related to the Cabibbo--Kobayashi--Maskawa (CKM)
matrix \cite{ckm}, and can be characterized by the unitarity triangle (UT) with its
three angles $\alpha$, $\beta$ and $\gamma$. This adventure will soon
be continued at the LHC (CERN), with its dedicated $B$-decay experiment 
LHCb.

In the $B$-physics landscape, an interesting probe of CP violation is also offered 
by $B^0_d\to D^+D^-$. As can be seen in Fig.~\ref{fig:BDD-diag}, this decay originates 
from $\bar b\to \bar c c\bar d$ quark-level processes, and receives contributions
both from a colour-allowed tree-diagram-like topology and from penguin diagrams. 
In analogy to the prominent $B^0_d\to\pi^+\pi^-$ decay, the latter contributions
lead to complications of the theoretical interpretation of the CP-violating observables.
However, the penguin effects can fortunately be controlled by means of the 
$B^0_s\to D^+_sD^-_s$ channel \cite{RF-BDD}, which is related to 
$B^0_d\to D^+D^-$ through an interchange of all down and strange quarks,
as can also be seen in Fig.~\ref{fig:BDD-diag}. Because of this feature, the 
$U$-spin flavour symmetry of strong interactions allows us to derive relations 
between non-perturbative hadronic parameters,\footnote{The $U$-spin
flavour symmetry connects strange and down quarks in the same way through 
$SU(2)$ transformations as the isopsin symmetry connects the up and down quarks.} 
so that the measurement of CP violation in $B^0_d\to D^+D^-$ can be converted into 
CP-violating weak phases. In comparison with conventional flavour-symmetry strategies 
\cite{GHLR}, the advantage of the $U$-spin method is that no additional dynamical 
assumptions are needed, and that also electroweak (EW) penguin contributions 
are automatically included. 

The key observables are the CP-averaged branching ratios as well as the direct 
and mixing-induced CP asymmetries ${\cal A}_{\rm CP}^{\rm dir}(B_d\to D^+D^-)$ 
and ${\cal A}_{\rm CP}^{\rm mix}(B_d\to D^+D^-)$, respectively, which enter
the following time-dependent rate asymmetry \cite{kitz}:
\begin{eqnarray}
\lefteqn{{\cal A}_{\rm CP}(B_d(t)\to D^+D^-)\equiv\frac{\Gamma(B^0_d(t)\to  D^+D^-)-
\Gamma(\bar B^0_d(t)\to  D^+D^-)}{\Gamma(B^0_d(t)\to  D^+D^-)+
\Gamma(\bar B^0_d(t)\to  D^+D^-)}}\nonumber\\
&&={\cal A}_{\rm CP}^{\rm dir}(B_d\to  D^+D^-)\,\cos(\Delta M_d t)+
{\cal A}_{\rm CP}^{\rm mix}(B_d\to D^+D^-)\,\sin(\Delta M_d t),\label{rate-asym}
\end{eqnarray}
where $\Delta M_d$ is the mass difference of the $B_d$ mass eigenstates.
The Belle collaboration has recently reported evidence for CP violation in
$B^0_d\to D^+D^-$, which could not be confirmed by BaBar. The current
status reads as follows:
\begin{equation}\label{ACP-dir-BDD-ex}
{\cal A}_{\rm CP}^{\rm dir}(B_d\to D^+D^-)=\left\{
\begin{array}{ll}
+0.11\pm0.22\pm0.07 & \mbox{(BaBar \cite{BaBar-BDD})}\\
-0.91\pm0.23\pm0.06 & \mbox{(Belle \cite{Belle-BDD})}
\end{array}\right.
\end{equation}
\begin{equation}\label{ACP-mix-BDD-ex}
{\cal A}_{\rm CP}^{\rm dir}(B_d\to D^+D^-)=\left\{
\begin{array}{ll}
+0.54\pm0.34\pm0.06 & \mbox{(BaBar \cite{BaBar-BDD})}\\
+1.13\pm0.37\pm0.09 & \mbox{(Belle \cite{Belle-BDD});}
\end{array}\right.
\end{equation}
the Heavy Flavour Averaging Group (HFAG) gives the following averages
\cite{HFAG}:
\begin{equation}\label{CP-HFAG}
{\cal A}_{\rm CP}^{\rm dir}(B_d\to D^+D^-)=-0.37 \pm 0.17, \quad
{\cal A}_{\rm CP}^{\rm mix}(B_d\to D^+D^-)=0.75 \pm 0.26,
\end{equation}
which have to be taken with great care in view of the inconsistency between the
BaBar and Belle measurements. Concerning the CP-averaged branching ratio, 
we have
\begin{equation}\label{BDD-BR}
\mbox{BR}(B_d\to D_d^+D_d^-)=\left\{
\begin{array}{ll}
(2.8\pm0.4\pm0.5)\times10^{-4}& \mbox{(BaBar \cite{BaBar-BDD-BR})}\\
(1.97\pm0.20\pm0.20)\times10^{-4} & \mbox{(Belle \cite{Belle-BDD}),}
\end{array}\right.
\end{equation}
yielding the average of $\mbox{BR}(B_d\to D_d^+D_d^-)=
(2.11\pm0.26)\times10^{-4}$. Thanks to the updated Belle result, this number is
now about $1.6\,\sigma$ lower than the HFAG value of 
$\mbox{BR}(B_d\to D_d^+D_d^-)=(3.0\pm0.5)\times10^{-4}$ \cite{HFAG}.
The CDF collaboration has recently observed the first signals of the 
$B^0_s\to D_s^+D_s^-$ decay \cite{CDF-BsDsDs}, which correspond to the 
CP-averaged branching ratio 
\begin{equation}\label{BsDsDs-BR}
\mbox{BR}(B_s\to D_s^+D_s^-)=(1.09\pm0.27\pm0.47)\%.
\end{equation}
Performing a run on the $\Upsilon(5S)$ resonance, also the Belle collaboration has
recently obtained an upper bound of $6.7\%$ (90\% C.L.) for this branching ratio
\cite{Belle-Y5S}. Moreover, the D0 collaboration has performed a first
analysis of the combined $B_s\to D_s^{(*)}D_s^{(*)}$ branching ratio, with 
the result of $\mbox{BR}(B_s\to D_s^{(*)}D_s^{(*)})=(3.9^{+1.9+1.6}_{-1.7-1.5})\%$
\cite{D0-BsDsDs}.

Although the current experimental picture is still in an early stage, it raises
several questions, which are further motivated by the quickly approaching
start of the LHC:
\begin{itemize}
\item What is the allowed SM region for the CP violation in $B^0_d\to D^+D^-$?
\item What are the most promising strategies for the extraction of weak phases?
\item What is the interplay with other measurements of CP violation and the 
search for new physics (NP)?
\end{itemize}
These items are the central target of this paper. It is organized as follows:
in Section~\ref{sec:obs-space}, we explore the parameter space of the
CP-violating $B^0_d\to D^+D^-$ asymmetries, taking also the constraints from
$B^0_s\to D^+_sD^-_s$ and similar modes into account, and perform a 
theoretical estimate of the corresponding observables in Section~\ref{sec:esti}. 
In Section~\ref{sec:extr}, we discuss the extraction of CP-violating phases from the 
$B^0_d\to D^+D^-$ and $B^0_s\to D^+_sD^-_s$ decays, while the interplay with other 
CP probes is discussed in Section~\ref{sec:inter}. Finally, we summarize our 
conclusions in Section~\ref{sec:concl}.

\begin{figure}
\centerline{
 \includegraphics[width=7.2truecm]{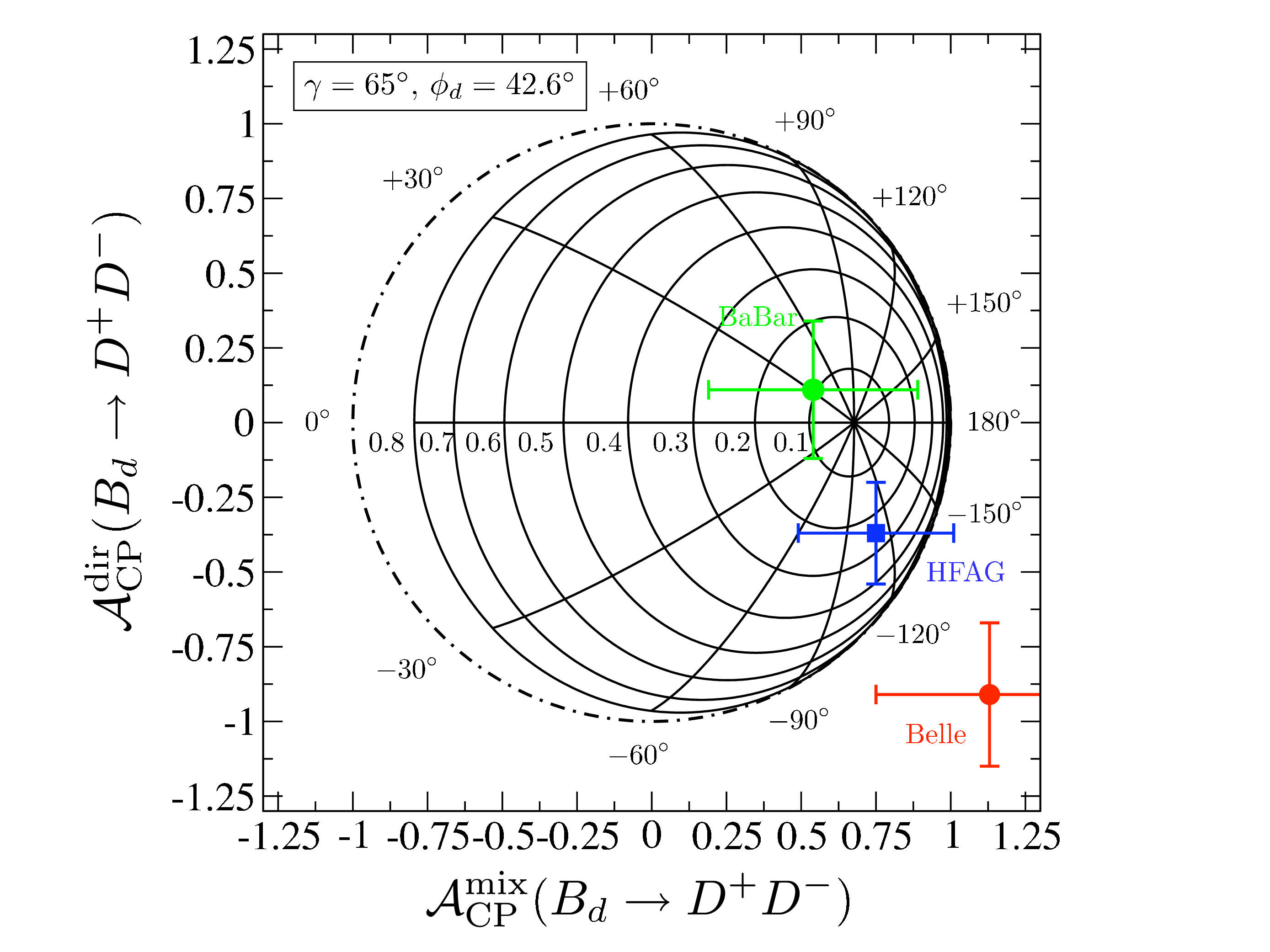}
 \hspace*{0.5truecm}
 \includegraphics[width=7.2truecm]{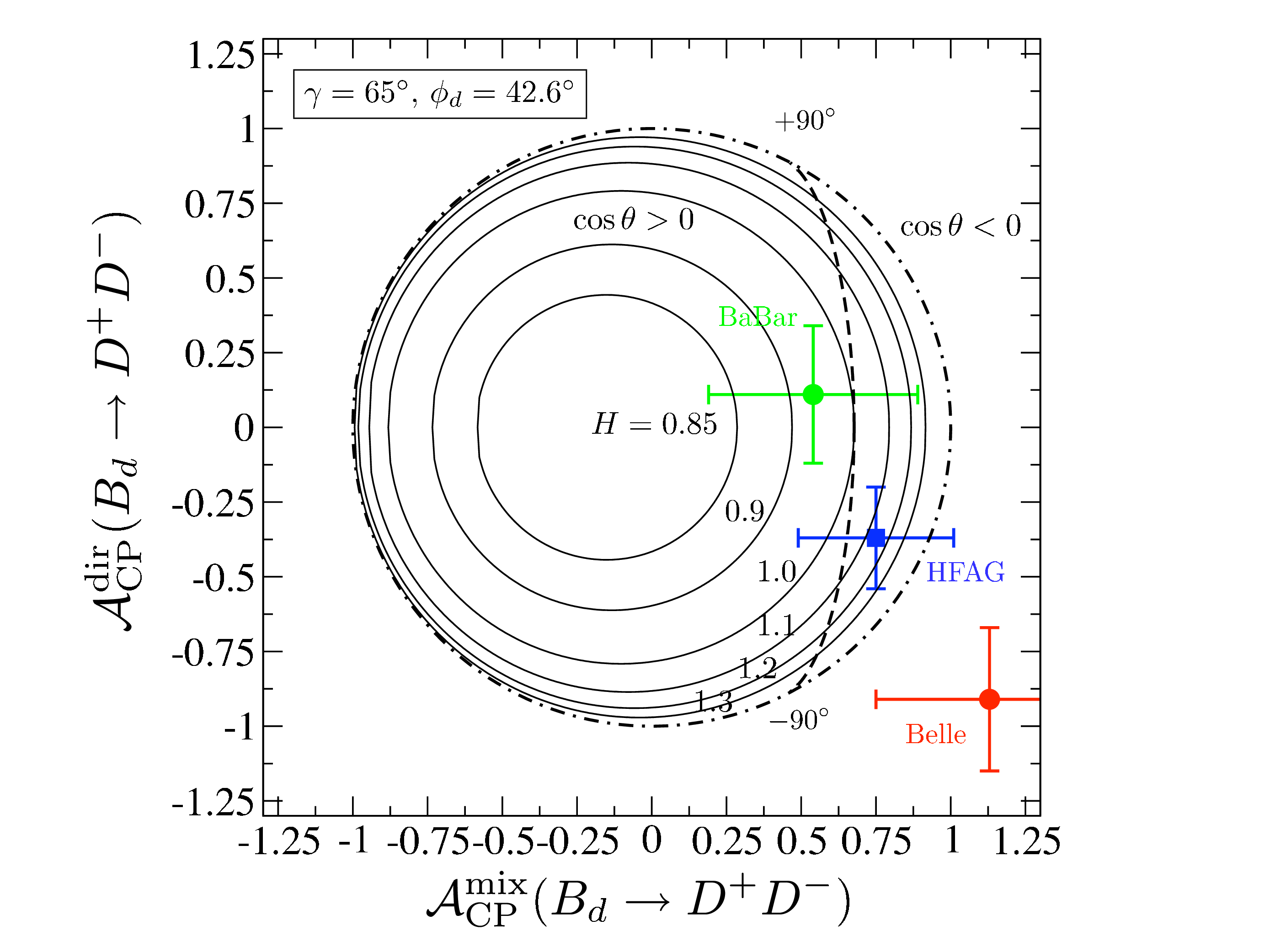} 
 }
\caption{The situation in the 
${\cal A}_{\rm CP}^{\rm mix}(B_d\to D^+D^-)$--${\cal A}_{\rm CP}^{\rm dir}
(B_d\to D^+D^-)$ plane. Left panel: contours following from the general SM
parametrization; right panel: constraints following from a measurement of
the quantity $H$.}\label{fig:AM-AD-gam65}
\end{figure}

\boldmath
\section{CP Violation in $B_d^0\to D^+D^-$}\label{sec:obs-space}
\unboldmath
\subsection{Standard Model Expressions}
In the SM, we may write the $B_d^0\to D^+D^-$ decay amplitude as follows 
\cite{RF-BDD}:
\begin{equation}\label{BDD-ampl}
A(B^0_d\to D^+D^-)=-\lambda{\cal A}\left[1-a e^{i\theta}e^{i\gamma}\right],
\end{equation}
where $\lambda$ is the well-known Wolfenstein parameter of the CKM matrix 
\cite{wolf}, ${\cal A}$ denotes a CP-conserving strong amplitude 
that is governed by the tree contributions, while the CP-consering hadronic 
parameter $a e^{i\theta}$ measures -- sloppily speaking -- the ratio of 
penguin to tree amplitudes. Applying the well-known formalism to 
calculate the CP-violating observables that are provided by the time-dependent
rate asymmetry in (\ref{rate-asym}), we obtain the following expressions:
\begin{eqnarray}
{\cal A}_{\rm CP}^{\rm dir}(B_d\to D^+D^-)&=&
\frac{2a\sin\theta\sin\gamma}{1-2a\cos\theta\cos\gamma+a^2}\label{Adir-expr}\\
{\cal A}_{\rm CP}^{\rm mix}(B_d\to D^+D^-)&=&
\frac{\sin\phi_d-2a\cos\theta\sin(\phi_d+\gamma)+a^2\sin(\phi_d+2\gamma)}{1-
2a\cos\theta\cos\gamma+a^2},\label{Amix-expr}
\end{eqnarray}
where $\phi_d$ denotes the $B^0_d$--$\bar B^0_d$ mixing phase, which takes
the value of $2\beta$ in the SM. This quantity has been measured at the $B$ factories 
with the help of the ``golden" decay $B^0_d\to J/\psi K_{\rm S}$ and similar modes, 
including $B_d\to J/\psi K^*$ and $B_d \to D^*D^*K_{\rm S}$ channels to resolve a 
twofold ambiguity, as follows \cite{HFAG}:
\begin{equation}\label{phid-exp}
\phi_d=(42.6\pm2)^\circ.
\end{equation} 
Concerning the angle $\gamma$, the SM fits of the UT obtained 
by the UTfit and CKMfitter collaborations \cite{UTfit,CKMfitter} yield 
$\gamma=(64.6 \pm 4.2)^\circ$ and $\gamma=(59.0^{+9.2}_{-3.7})^\circ$, 
respectively. A recent analysis of the $U$-spin-related $B_d\to\pi^+\pi^-$ and
$B_s\to K^+K^-$ transitions finds $\gamma=(66.6^{+4.3+4.0}_{-5.0-3.0})^\circ$ 
\cite{RF-Bpipi-07}, in excellent agreement with these fits. A similar picture
emerges also from other recent $\gamma$ determinations from $B\to\pi\pi,\pi K$
decays \cite{FRS,GR-07}. Thanks to the LHCb experiment \cite{LHCb}, our 
knowledge of $\gamma$ will soon improve dramatically, also since very accurate 
``reference" determinations through pure tree decays will become available. 
In the limit of $a\to 0$, expression (\ref{Amix-expr}) would allow a straighforward 
extraction of $\sin\phi_d$. However, these penguin effects cannot simply be neglected
and require further work. 

For the following discussion, we shall assume $\gamma=65^\circ$ and
$\phi_d=42.6^\circ$. Using (\ref{Adir-expr}) and (\ref{Amix-expr}), we can then
calculate contours in the 
${\cal A}_{\rm CP}^{\rm mix}(B_d\to D^+D^-)$--${\cal A}_{\rm CP}^{\rm dir}
(B_d\to D^+D^-)$ plane for given values of $a$ and $\theta$, which are
theoretically exact in the SM. The resulting picture is shown in the left 
panel of Fig.~\ref{fig:AM-AD-gam65} (for its $B^0_d\to\pi^+\pi^-$ counterpart,
see Ref.~\cite{FlMa}). There we have also included the
experimental BaBar and Belle results, as well as the HFAG average; the 
dot-dashed circle defines the outer boundary in this observable space that 
follows from the general relation
\begin{equation}
\left[{\cal A}_{\rm CP}^{\rm dir}(B_d\to D^+D^-)\right]^2+
\left[{\cal A}_{\rm CP}^{\rm mix}(B_d\to D^+D^-)\right]^2\leq1.
\end{equation}
Fig.~\ref{fig:AM-AD-gam65} shows that the Belle result lies outside
of the physical region, in contrast to the BaBar measurement and the 
HFAG average. The contours of that figure allow us to read off the
corresponding values of $a$ and $\theta$ straightforwardly.

\boldmath
\subsection{Constraints from $B^0_s\to D^+_sD^-_s$}
\unboldmath
We now go one step further by using the information that is offered by the
$B^0_s\to D^+_sD^-_s$ decay. In analogy to (\ref{BDD-ampl}),  its
SM amplitude can be written as follows:
\begin{equation}\label{ABsDsDs-ampl}
A(B^0_s\to D^+_sD^-_s)=\left(1-\frac{\lambda^2}{2}\right){\cal A}'
\left[1+\epsilon a' e^{i\theta'}e^{i\gamma}\right],
\end{equation}
where 
\begin{equation}
\epsilon\equiv\frac{\lambda^2}{1-\lambda^2}=0.05.
\end{equation}
Following Ref.~\cite{RF-BDD}, we introduce 
\begin{eqnarray}
H&\equiv&\frac{1}{\epsilon}\,\left|\frac{{\cal A}'}{{\cal A}}\right|^2
\left[\frac{M_{B_d}}{M_{B_s}}\,\frac{\Phi(M_{D_s}/M_{B_s},M_{D_s}/M_{B_s})}{
\Phi(M_{D_d}/M_{B_d},M_{D_d}/M_{B_d})}\,\frac{\tau_{B_s}}{\tau_{B_d}}\right]
\left[\frac{\mbox{BR}(B_d\to D^+D^-)}{\mbox{BR}(B_s\to D_s^+D_s^-)}\right]\nonumber\\
&&\hspace*{2.5truecm}=\frac{1-2a\cos\theta\cos\gamma+a^2}{1+
2\epsilon a'\cos\theta'\cos\gamma+\epsilon^2 a'^2},\label{H-def}
\end{eqnarray}
where 
\begin{equation}
\Phi(x,y)\equiv\sqrt{\left[1-(x+y)^2\right]\left[1-(x-y)^2\right]}
\end{equation}
is the well-known $B\to PP$ phase-space function, and the 
$\tau_{B_{d,s}}$ are the $B_{d,s}$ lifetimes. Applying the $U$-spin flavour
symmetry, we obtain the relations
\begin{equation}\label{U-spin-rel1}
a'=a, \quad \theta'=\theta.
\end{equation}
Thanks to the $\epsilon$ suppression in (\ref{H-def}), the impact of $U$-spin-breaking
corrections to (\ref{U-spin-rel1}) is marginal for $H$. In the case of $|{\cal A}'/{\cal A}|$, 
ratios of $U$-spin-breaking decay constants and form factors enter. If we apply the
``factorization'' approximation, we obtain
\begin{equation}\label{SU3-breakBDD}
\left|\frac{{\cal A'}}{{\cal A}}\right|_{\rm fact}=
\frac{(M_{B_s}-M_{D_s})\,\sqrt{M_{B_s}M_{D_s}}\,(w_s+1)}{(M_{B_d}-M_{D_d})
\,\sqrt{M_{B_d}M_{D_d}}\,(w_d+1)}\frac{f_{D_s}\,\xi_s(w_s)}{f_{D_d}\,
\xi_d(w_d)}\,,
\end{equation}
where the restrictions form the heavy-quark effective theory for the 
$B_q\to D_q$ form factors have been taken into account by introducing 
appropriate Isgur--Wise functions $\xi_q(w_q)$ with $w_q=M_{B_q}/(2M_{D_q})$ 
\cite{neu-ste}. Studies of the light-quark dependence of the Isgur--Wise
function were performed within heavy-meson chiral perturbation theory, 
indicating an enhancement of $\xi_s/\xi_d$ at the level of $5\%$ 
\cite{HMChiPT1}. Applying the same formalism to $f_{D_s}/f_{D_d}$ gives values
at the 1.2 level \cite{HMChiPT2}, which is in accordance with the recent
measurement by the CLEO collaboration \cite{cleo}:
\begin{equation}\label{cleo}
\frac{f_{D_s}}{f_{D_d}}=1.23\pm0.11\pm0.04,
\end{equation}
as well as with lattice QCD calculations, as summarized in Ref.~\cite{onogi}.
Using heavy-meson chiral perturbation theory and the $1/N_{\rm C}$ expansion,
non-factorizable $SU(3)$-breaking corrections were found at the level of
a few percent in Ref.~\cite{RF-BDD-93}. The CDF result in (\ref{BsDsDs-BR})
and the average of (\ref{BDD-BR}) yield then, with the CLEO measurement
in (\ref{cleo}), the following numbers:
\begin{equation}\label{H-exp}
H= 0.59 \pm 0.31 \quad (0.84\pm0.45),
\end{equation}
where we have added the errors in quadrature, and have also given
the result corresponding to the HFAG value of $\mbox{BR}(B_d\to D^+D^-)$
in parentheses. The general expression for $H$ in (\ref{H-def}) implies a
lower bound \cite{FR2}, which is given by
\begin{equation}
H\geq\left[1-2\epsilon\cos^2\gamma+{\cal O}(\epsilon^2)\right]\sin^2\gamma
\stackrel{\gamma=65^\circ}{\longrightarrow}0.81.
\end{equation}
Consequently, the rather low central value of (\ref{H-exp}), which is essentially
due to the new Belle result \cite{Belle-BDD}, is disfavoured by the 
experimental information on $\gamma$. 

If we replace the $s$ spectator quark of the $B^0_s\to D^+_s D^-_s$ decay through 
a $d$ quark, we obtain the $B^0_d\to D_s^+ D^-$ process. Whereas
the $B_{d(s)}\to D^+_{d(s)}D^-_{d(s)}$ system receives contributions from tree and 
penguin as well as exchange ($E$) and penguin annihilation ($PA$) topologies
(the latter are not shown in Fig.~\ref{fig:BDD-diag}), the  $B^0_d\to D_s^+ D^-$
channel and its $U$-spin partner $B^0_s\to D^+D^-_s$ receive only tree and
penguin contributions. Consequently, if we use the $SU(3)$ flavour
symmetry and assume that the exchange and penguin annihilation topologies
play a minor r\^ole, we may replace $B^0_s\to D^+_s D^-_s$ in the determination 
of $H$ through $B^0_d\to D_s^+ D^-$ \cite{datta}.\footnote{This is analogous
to the replacement of $B^0_s\to K^+K^-$ through $B^0_d\to\pi^-K^+$
\cite{RF-BsKK}.} Expression (\ref{H-def}) is then modified as follows:
\begin{equation}\label{H-1}
H\approx\frac{1}{\epsilon}\,\left(\frac{f_{D_s}}{f_{D_d}}\right)^2
\left[\frac{\Phi(M_{D_s}/M_{B_d},M_{D_d}/M_{B_d})}{
\Phi(M_{D_d}/M_{B_d},M_{D_d}/M_{B_d})}\right]
\left[\frac{\mbox{BR}(B_d\to D^+D^-)}{\mbox{BR}(B_d\to D_s^\pm D^\mp)}
\right].
\end{equation}
The importance of the $E+PA$ amplitude can actually be
probed through the $U$-spin related $B_{d(s)}\to D^+_{s(d)}D^-_{s(d)}$ decays.
The current experimental situation can be summarized as follows:
\begin{equation}
\mbox{BR}(B_d\to D_s^\pm D^\mp)=\left\{
\begin{array}{ll}
(6.4\pm1.3\pm1.0)\times10^{-3} & \mbox{(BaBar \cite{BaBar-BDsD})}\\
(7.5\pm0.2\pm0.8\pm0.8)\times10^{-3} & \mbox{(Belle \cite{Belle-BDsD}),}
\end{array}\right.
\end{equation}
yielding the average of $\mbox{BR}(B_d\to D_s^\pm D^\mp)=(7.1\pm0.9)\times10^{-3}$;
Belle reported also the upper limit of
$\mbox{BR}(B_d\to D_s^+ D_s^-)<3.6\times 10^{-5}$ (90\% C.L.) \cite{Belle-BDsD}.
Expression (\ref{H-1}) gives then 
\begin{equation}
H=  0.85 \pm 0.19  \quad (1.22 \pm 0.31),
\end{equation}
where the notation is as in (\ref{H-exp}).

Let us now investigate the constraints on the 
${\cal A}_{\rm CP}^{\rm mix}(B_d\to D^+D^-)$--${\cal A}_{\rm CP}^{\rm dir}
(B_d\to D^+D^-)$ plane that follow from $H$. If we use (\ref{H-def}) with
(\ref{U-spin-rel1}), we may eliminate the strong phase $\theta$ in (\ref{Adir-expr}) 
and (\ref{Amix-expr}) with the help of
\begin{equation}
\cos\theta=\frac{1-H+(1-\epsilon^2H)a^2}{2a(1+\epsilon H)\cos\gamma}, \quad
\sin\theta=\pm\sqrt{1-\cos^2\theta}.
\end{equation}
If we then keep $a$ as a free parameter, we arrive at the situation shown in the 
right panel of Fig.~\ref{fig:AM-AD-gam65}, where the dashed line separates the 
regions with $\cos\theta>0$ and $\cos\theta<0$. In the factorization approximation, 
we expect a negative value of $\cos\theta$. Although non-factorizable effects could
generate a large value of $\theta$, we do not expect that $\cos\theta$ changes its 
sign. This feature is in fact observed for other non-leptonic $B$-meson decays, such 
as the $B^0_d\to\pi^+\pi^-$, $B^0_d\to\pi^- K^+$ system \cite{RF-Bpipi-07}. With
$\gamma\sim65^\circ$, which corresponds to $\cos\gamma>0$, the expression
in (\ref{H-def}) implies then $H>1$. In the right panel of Fig.~\ref{fig:AM-AD-gam65}, 
this leaves us with the banana-shaped region in the 
${\cal A}_{\rm CP}^{\rm mix}(B_d\to D^+D^-)$--${\cal A}_{\rm CP}^{\rm dir}
(B_d\to D^+D^-)$ plane. Interestingly, the central value of the HFAG average
falls well into this region, whereas the central value of the BaBar result would
require a positive value of $\cos\theta$. Although the current errors are too
large to draw definite conclusions, this exercise illustrates the usefulness
of the plots in observable space to monitor the experimental picture. Since
the $B_s$ input for the determination of $H$ is just the CP-averaged
$B_s\to D_s^+D_s^-$ branching ratio, this measurement would also be
interesting for an $e^+e^-$ (super-)$B$ factory operating at the $\Upsilon(5S)$
resonance \cite{Belle-Y5S,super-B-Y5S}.

\begin{figure}
\centerline{
 \includegraphics[width=7.8truecm]{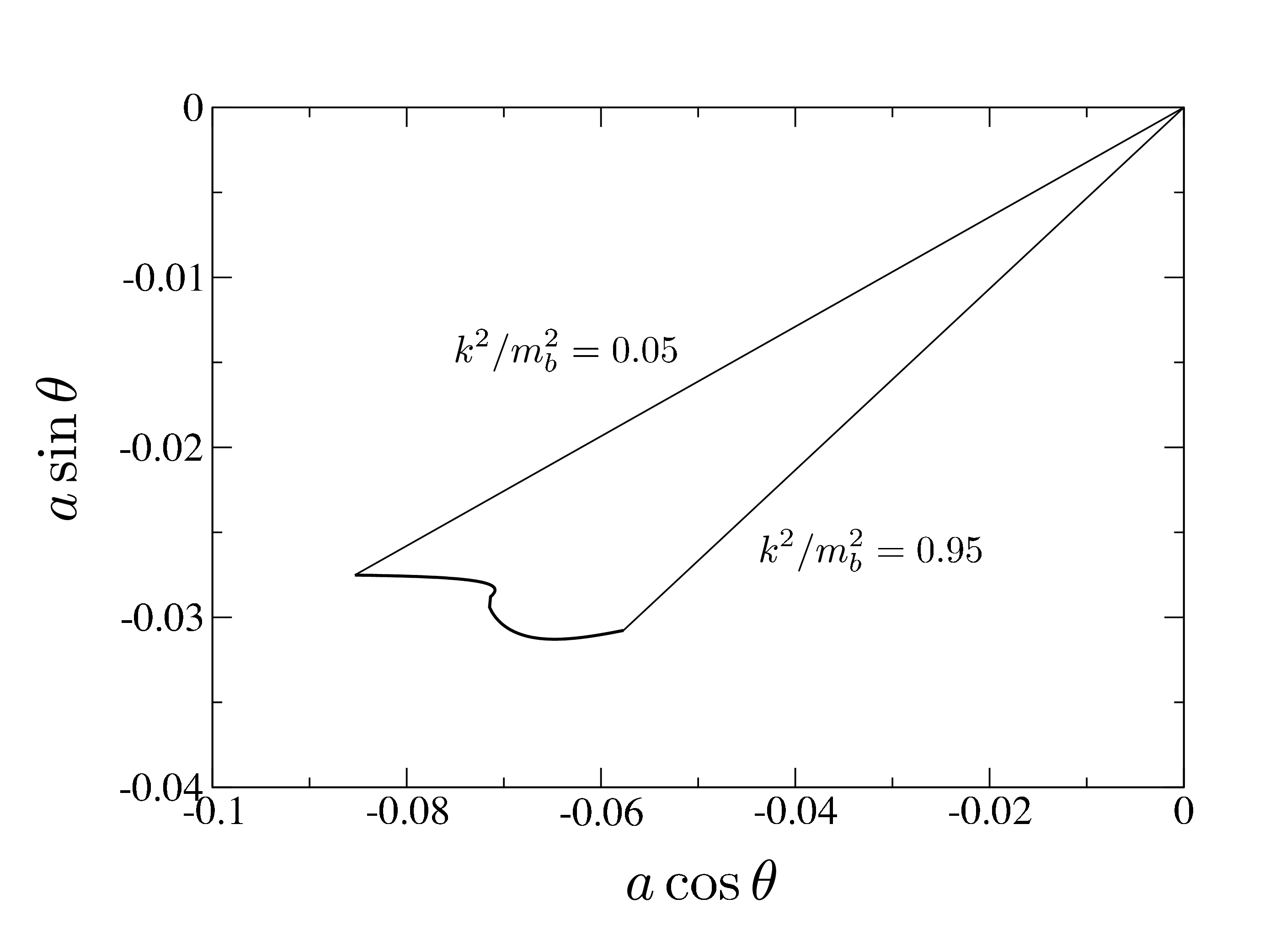} 
 \hspace*{0.5truecm} 
 \includegraphics[width=6.9truecm]{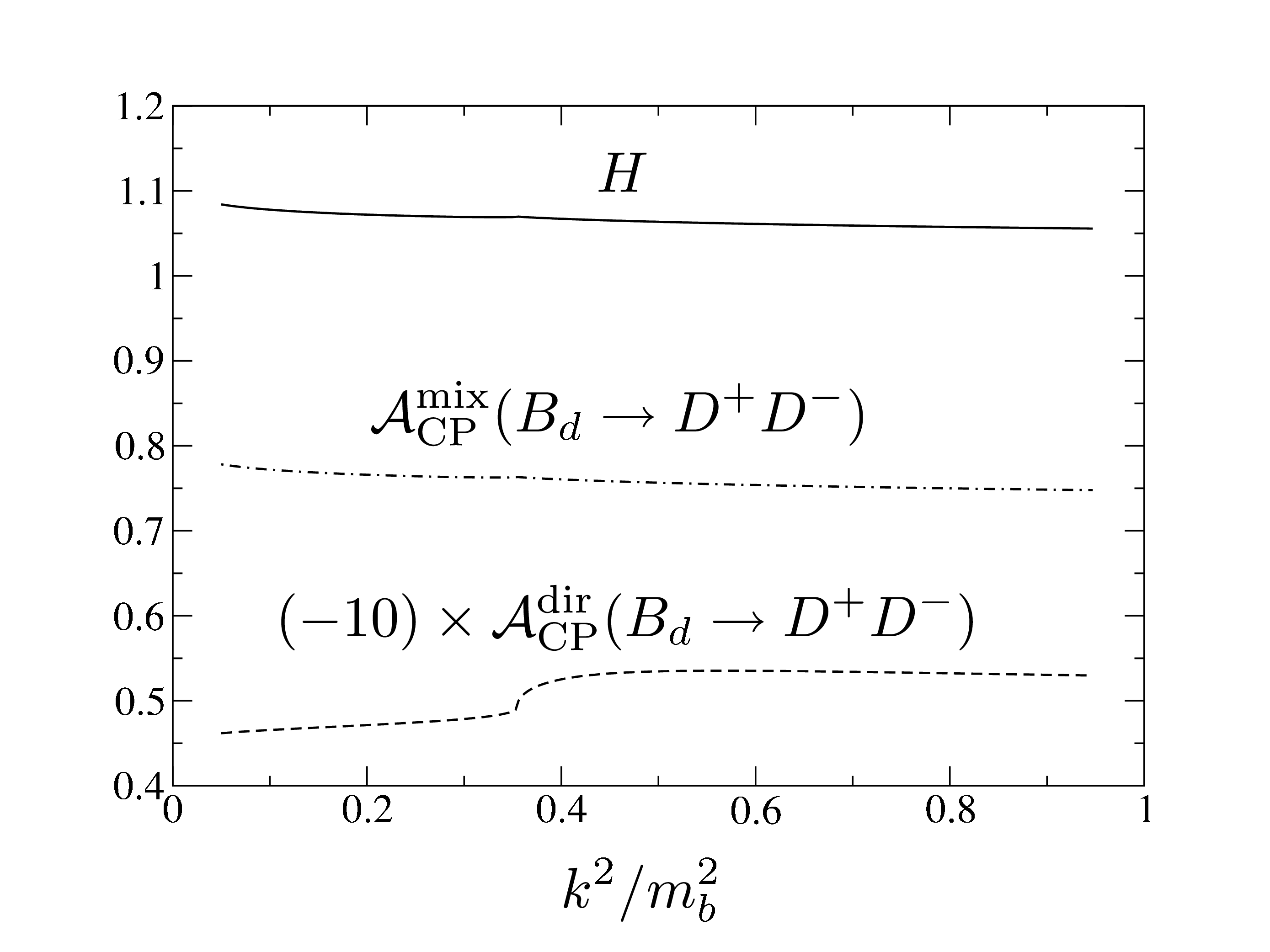}
 }
 \vspace*{-0.2truecm}
\caption{Theoretical estimates of the hadronic parameter $a e^{i\theta}$
(left panel), and the $B_{d(s)}\to D_{d(s)}^+D_{d(s)}^-$ observables 
(right panel) for $\gamma=65^\circ$, $\phi_d=42.6^\circ$ and
$R_b=0.45$.}\label{fig:calc}
\end{figure}

\section{Theoretical Estimates}\label{sec:esti}
In order to analyze the $B^0_d\to D^+D^-$ decay theoretically, we ``integrate out" 
the heavy degrees of freedom, i.e.\ the $W$ boson and top quark in 
Fig.~\ref{fig:BDD-diag}, and use an appropriate low-energy effective Hamiltonian, 
which takes the following form \cite{BBL-rev}:
\begin{equation}\label{heff}
{\cal H}_{{\rm eff}} = \frac{G_{\rm F}}{\sqrt{2}}\left[ 
\lambda_u^{(d)}\sum_{k=1}^2C_k(\mu) Q_k^{ud} +\lambda_c^{(d)}
\sum_{k=1}^2C_k(\mu) Q_k^{cd} -\lambda_t^{(d)} \sum^{10}_{k=3} 
C_k(\mu) Q_k^d \right].
\end{equation}
Here the $\lambda_j^{(d)}\equiv V_{jd}V_{jb}^\ast$ denote CKM factors, 
$Q_1^{jd}$ and $Q_2^{jd}$ ($j\in\{u,c\}$) are the usual 
current--current operators, $Q_3^d, \ldots ,Q_6^d$ and $Q_7^d, \ldots ,
Q_{10}^d$ denote the QCD and EW penguin operators, respectively, 
and $\mu={\cal O}(m_b)$ is a renormalization scale. If we apply the 
Bander--Silverman--Soni mechanism \cite{bss} as well as the formalism 
developed in Ref.~\cite{pen-calc}, we obtain the following estimate:
\begin{equation}\label{a-approx}
a e^{i\theta}\approx R_b\left[\frac{{\cal A}_t+
{\cal A}_u}{{\cal A}_{\rm T}+{\cal A}_t+{\cal A}_c}\right],
\end{equation}
where $R_b\propto |V_{ub}/V_{cb}|$ is the corresponding side of the UT, and
\begin{eqnarray}
{\cal A}_{\rm T}&=&\frac{1}{3}\,\overline{C}_1+\overline{C}_2\label{AT}\\
{\cal A}_t&=&\frac{1}{3}\left[\overline{C}_3+\overline{C}_9+
\chi_D\left(\overline{C}_5+\overline{C}_7\right)\right]+\overline{C}_4+
\overline{C}_{10}+\chi_D\left(\overline{C}_6+\overline{C}_8\right)\label{A0}\\
{\cal A}_j&=&\frac{\alpha_s}{9\pi}\left[\frac{10}{9}-G(m_j,k,m_b)\right]
\left[\overline{C}_2+\frac{1}{3}\frac{\alpha}{\alpha_s}\left(3\,\overline{C}_1
+\overline{C}_2\right)\right]\left(1+\chi_D\right),\label{Aq}
\end{eqnarray}
with $j\in\{u,c\}$. The $\overline{C}_k$ refer to $\mu=m_b$ 
and denote the next-to-leading order scheme-independent Wilson coefficient 
functions introduced in Ref.~\cite{Buras-NLO}. The quantity
\begin{equation}
\chi_D=\frac{2M_D^2}{(m_c+m_d)(m_b-m_c)}
\end{equation}
is due to the use of the equations of motion for the quark fields, whereas 
the function $G(m_j,k,m_b)$ originates from the one-loop penguin matrix 
elements of the current--current operators $Q_{1,2}^{jq}$ with internal 
$j$ quarks. It is given by 
\begin{equation}
G(m_j,k,m_b)=-\,4\int\limits_{0}^{1}\mbox{d}x\,x\,(1-x)\ln\left[\frac{m_j^{2}-
k^{2}\,x\,(1-x)}
{m_b^{2}}\right],
\end{equation}
where $m_j$ is the $j$-quark mass and $k$ denotes some average four-momentum 
of the virtual gluons and photons appearing in the penguin diagrams 
\cite{pen-calc}. In Fig.~\ref{fig:calc}, we show the corresponding results,
keeping $k^2$ as a free parameters. The sensitivity on $k^2$ is moderate,
and in the case of $H$ and the mixing-induced CP asymmetry even small.
It should be emphasized that these results, with $a\sim0.08$ and $\theta\sim 205^\circ$
yielding the observables $H\sim1.07$, 
${\cal A}_{\rm CP}^{\rm dir}(B_d\to D^+D^-)\sim-5\%$ and
${\cal A}_{\rm CP}^{\rm dir}(B_d\to D^+D^-)\sim76\%$, can only be considered 
as estimates. A similar analysis was also performed in Ref.~\cite{xing}; however,
in Eq.~(12) of that paper, a factor of $\xi$ is missing in front of $C_3$, and 10/3
should read 10/9.

It is instructive to compare (\ref{a-approx}) with the corresponding expression
for the penguin-to-tree ratio $d$ of the $B^0_d\to \pi^+\pi^-$ decay in 
Ref.~\cite{RF-BsKK}. We observe that $a$ is suppressed with respect to
$d$ by a factor of $R_b^2\sim0.2$. The value of $d\sim0.4$, as determined
from the $U$-spin analysis of the $B_d\to\pi^+\pi^-$, $B_s\to K^+K^-$ system 
\cite{RF-Bpipi-07}, points therefore also towards $a\sim 0.08$. However, the 
detailed dynamics of these decays is of course very different, so that values 
of $a$ at the $20\%$ level cannot be excluded.

\begin{figure}
\centerline{
 \includegraphics[width=7.8truecm]{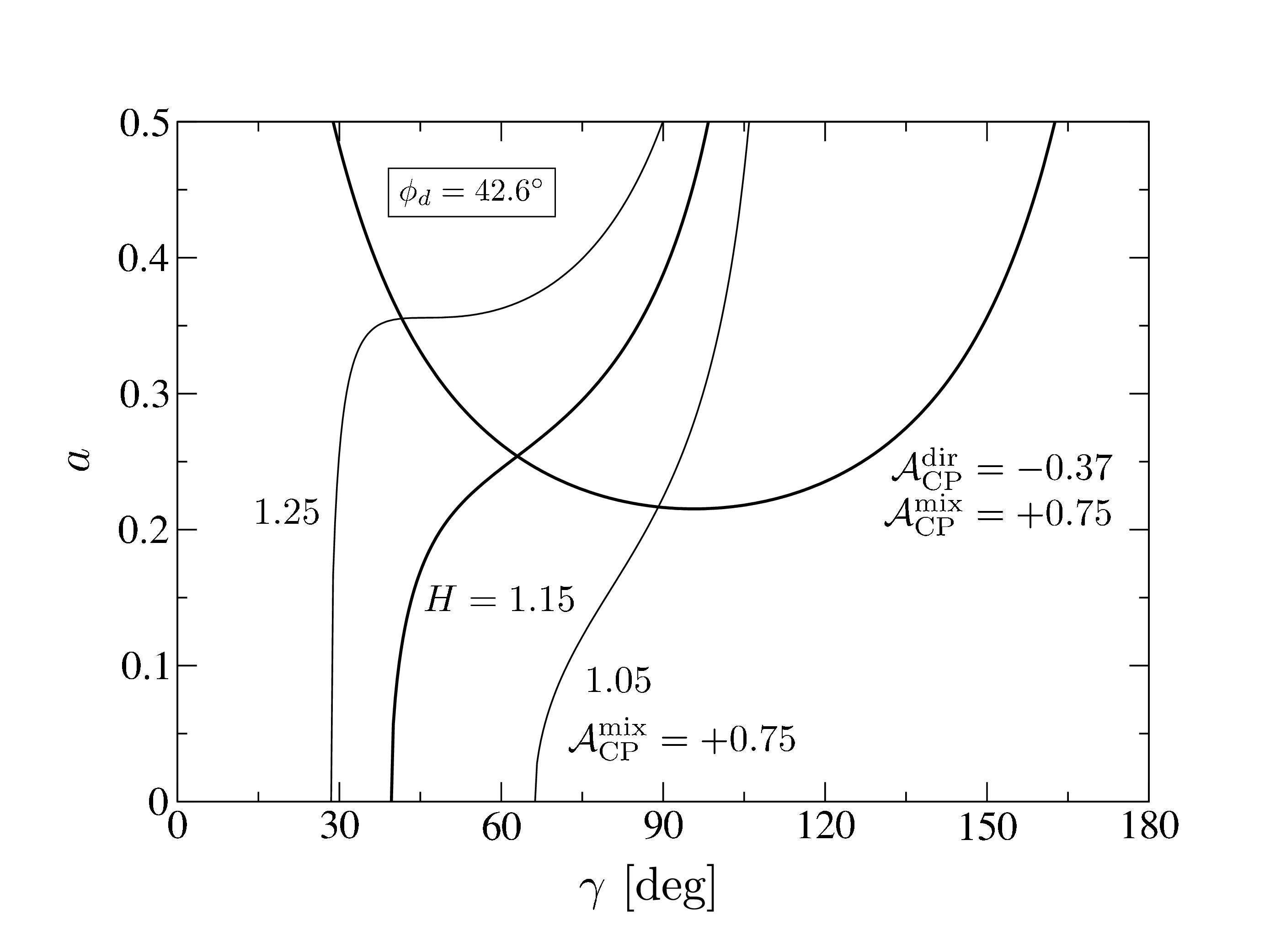}
 }
\caption{Illustration of the contours in the $\gamma$--$a$ plane for the
central values of the CP-violating $B_d\to D^+D^-$ asymmetries in 
(\ref{CP-HFAG}) and various values of the ratio $H$ of the 
CP-averaged $B_d\to D^+D^-$, $B_s\to D_s^+D_s^-$ branching 
ratios.}\label{fig:gam-a-cont}
\end{figure}

\boldmath
\section{Extractions of CP-Violating Phases}\label{sec:extr}
\unboldmath
\boldmath
\subsection{Extraction of $\gamma$}
\unboldmath
As was pointed out in Ref.~\cite{RF-BDD}, we may combine (\ref{Adir-expr})
with (\ref{Amix-expr}) to eliminate the strong phase $\theta$, which allows us
to calculate $a$ as a function of $\gamma$. To this end, the $B^0_d$--$\bar B^0_d$
mixing phase $\phi_d$ is needed as an input. The corresponding contour
relies only on the SM structure of the $B^0_d\to D^+D^-$ decay amplitude
and is {\it theoretically clean}. A second curve of this kind can be fixed through
${\cal A}_{\rm CP}^{\rm mix}(B_d\to D^+D^-)$ and $H$ with the help of
the $U$-spin relations in (\ref{U-spin-rel1}). The advantage of the combination
of these observables is that they both depend on $\cos\theta^{(')}$. 
Because of the $\epsilon$ suppression of the $a'$ terms in (\ref{H-def}), 
$U$-spin-breaking corrections to this relation have actually a very small impact,
so that the major non-factorizable $U$-spin-breaking effects enter through the 
determination of $H$. In Fig.~\ref{fig:gam-a-cont}, we illustrate this strategy for
the central values of the averages in (\ref{CP-HFAG}) and different values of
$H$. We see that $H=1.15$ would give a value of $\gamma=63^\circ$
with $a=0.25$ (and $\theta=249^\circ$). On the other hand, $H=1.05$ yields
$\gamma=89^\circ$ with $a=0.22$ and $\theta=244^\circ$, whereas $H=1.25$
results in $\gamma=42^\circ$, $a=0.35$ and $\theta=257^\circ$. Consequently, 
since a variation of $H=1.15\pm0.10$ gives the large range of 
$\gamma=(63^{+26}_{-21})^\circ$, the situation would not be favourable 
for the determination of this  UT angle. However, the 
hadronic parameter $a=0.25^{+0.10}_{-0.03}$ and -- in particular 
the strong phase $\theta=(249^{+8}_{-5})^\circ$ -- could be well determined,
but are of less interest. In the case of the $U$-spin-related $B_d\to \pi^+\pi^-$, 
$B_s\to K^+K^-$ decays, the current data result in a complementary 
situation, with a very favourable situation for the extraction of $\gamma$, and a 
less fortunate picture for the corresponding strong phase \cite{RF-Bpipi-07}. 
It will be interesting to follow the future evolution of the 
$B_{d(s)}\to D^+_{d(s)}D^-_{d(s)}$ data.

\boldmath
\subsection{Extraction of the $B^0_d$--$\bar B^0_d$ Mixing Phase}
\unboldmath
An alternative avenue for extracting information from the CP-violating
asymmetries of the $B^0_d\to D^+D^-$ decay arises if we use $\gamma$ 
as an input. By the time accurate measurements of
these CP asymmetries will become available we will also have a 
clear picture of this UT angle thanks to the precision measurements
that can be performed at LHCb \cite{LHCb}. For the 
following analysis, we assume a value of $\gamma=65^\circ$
(see the remarks after (\ref{phid-exp})).

\begin{figure}
\centerline{
 \includegraphics[width=7.8truecm]{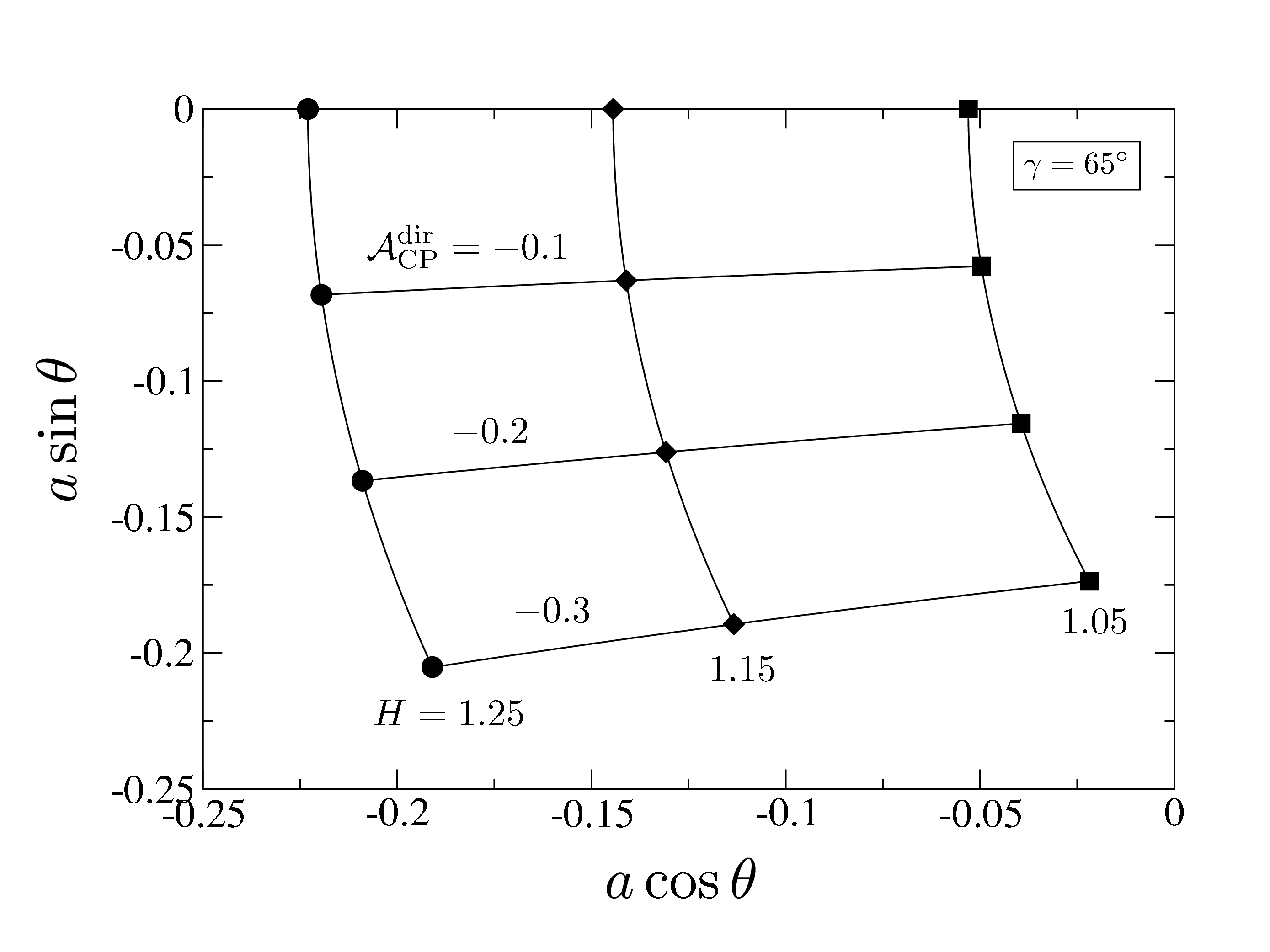}
 }
\caption{Determination of the hadronic parameter $ae^{i\theta}$ for given 
values of $H$ and ${\cal A}_{\rm CP}^{\rm dir}(B_d\to D^+D^-)$.}\label{fig:hadr-det}
\end{figure}

If the penguin effects could be neglected, the following simple situation would arise:
\begin{equation}
(\sin2\beta)_{D^+D^-}\equiv{\cal A}_{\rm CP}^{\rm mix}(B_d\to D^+D^-)
\stackrel{\rm no~pengs.}{\longrightarrow}\sin\phi_d\stackrel{\rm SM}{=}
\sin2\beta.
\end{equation}
The goal of the following discussion is to include the penguin effects in the
determination of $\sin\phi_d$. To this end, we first determine $a$ 
through the combination of ${\cal A}_{\rm CP}^{\rm dir}(B_d\to D^+D^-)$ 
and $H$ by means of the $U$-spin relation (\ref{U-spin-rel1}), which yields
\begin{equation}\label{a-det}
a=\sqrt{b-\sqrt{b^2-c}},
\end{equation}
where
\begin{eqnarray}
b N &=& 2\left[(1+\epsilon H)\sin\gamma\cos\gamma\right]^2+
(H-1)(1-\epsilon^2 H)\sin^2\gamma\nonumber\\
&&-\epsilon\left[(1+\epsilon)H{\cal A}_{\rm CP}^{\rm dir}(B_d\to D^+D^-)
\cos\gamma\right]^2
\end{eqnarray}
\begin{equation}
c N = \left[(H-1)\sin\gamma\right]^2+\left[(1+\epsilon)H
{\cal A}_{\rm CP}^{\rm dir}(B_d\to D^+D^-)\cos\gamma\right]^2,
\end{equation}
with
\begin{equation}
N=\left[(1-\epsilon^2 H)\sin\gamma\right]^2+
\left[\epsilon(1+\epsilon)H {\cal A}_{\rm CP}^{\rm dir}(B_d\to D^+D^-)
\cos\gamma\right]^2.
\end{equation}
In (\ref{a-det}), the sign in front of the inner square root could, in principle, 
be positive or negative. However, since the large values of $a$ corresponding 
to the $+$ sign are completely unrealistic, we have already written the $-$
sign. The strong phase $\theta$ follows then from
\begin{equation}\label{ctheta-det}
\cos\theta=\frac{1-H+(1-\epsilon^2H)a^2}{2(1+\epsilon H)a\cos\gamma}
\end{equation}
\begin{equation}
\sin\theta=\left[\frac{(1+\epsilon)(1+\epsilon a^2)}{2(1+\epsilon H)a\sin\gamma}\right]
H{\cal A}_{\rm CP}^{\rm dir}(B_d\to D^+D^-).
\end{equation}
In these expressions, the impact of the $\epsilon$ terms is tiny, but we have
kept them for completeness. In Fig.~\ref{fig:hadr-det}, we show the
resulting picture of the hadronic parameter $ae^{i\theta}$ in the complex 
plane for various values of $H$ and ${\cal A}_{\rm CP}^{\rm dir}(B_d\to D^+D^-)$, 
which should be compared with theoretical estimate shown in the left panel 
of Fig.~\ref{fig:calc}.

\begin{figure}
\centerline{
 \includegraphics[width=7.8truecm]{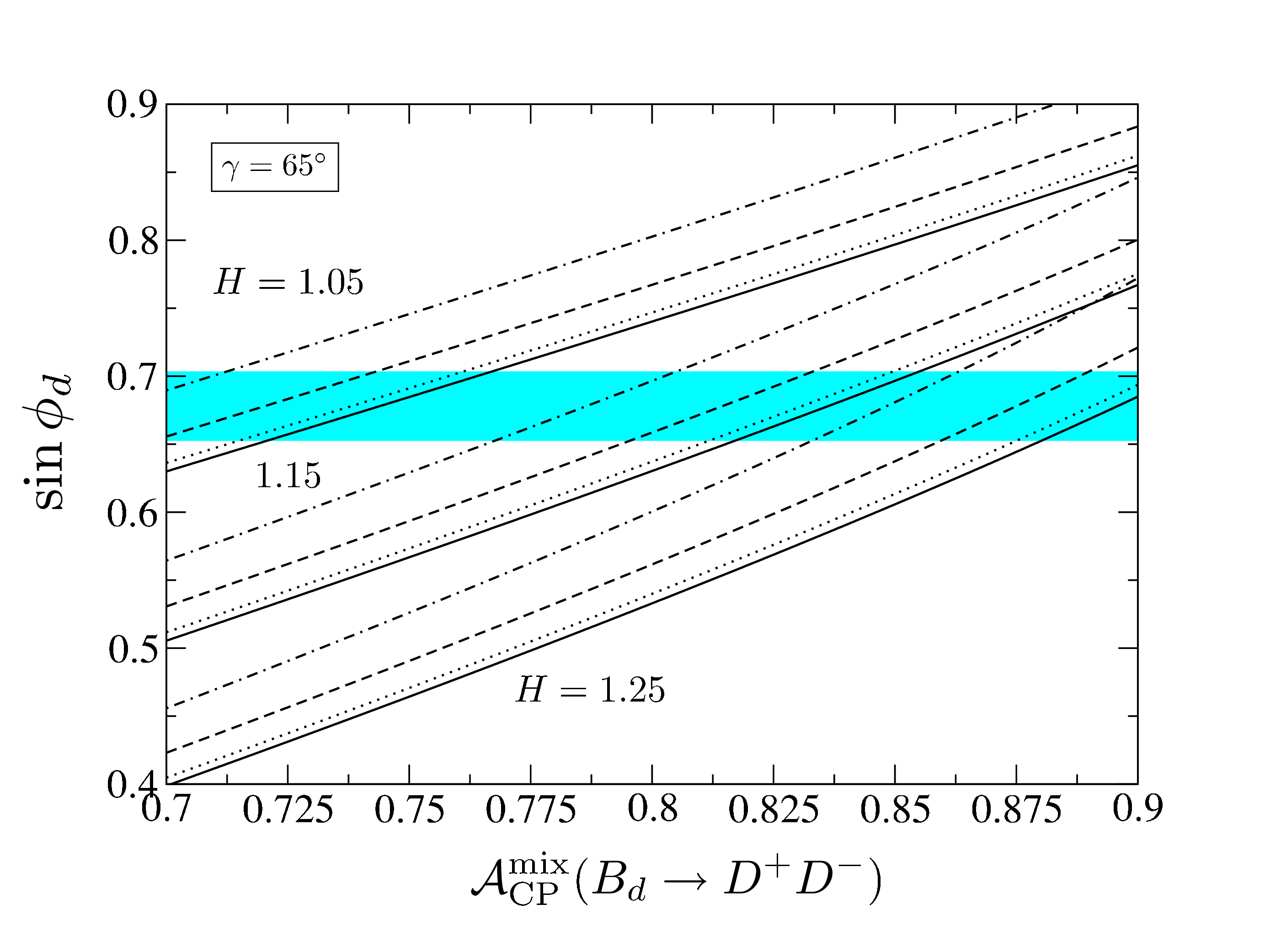}
 }
\caption{Correlation between ${\cal A}_{\rm CP}^{\rm mix}(B_d\to D^+D^-)$ 
and $\sin\phi_d$ for given values of $H$ and various values of 
${\cal A}_{\rm CP}^{\rm dir}(B_d\to D^+D^-)$: $0$ (solid), $\pm0.1$ (dotted),
$\pm0.2$ (dashed), $\pm0.3$ (dot-dashed). The shaded region corresponds
to the experimental value of $(\sin2\beta)_{\psi K_{\rm S}}$.}\label{fig:Amix-s2b}
\end{figure}

If we now use again (\ref{U-spin-rel1}) and eliminate $\cos\theta$ in 
(\ref{Amix-expr}) through (\ref{ctheta-det}), we obtain
\begin{equation}
A\sin\phi_d+B\cos\phi_d=C,
\end{equation}
where
\begin{eqnarray}
A&=&\left[H-2a^2\sin^2\gamma+\epsilon H\left\{1+\left(1-2\sin^2\gamma+\epsilon
\right)a^2\right\}\right]\cos\gamma\\
B&=&\left[H-1+a^2\cos2\gamma+\epsilon H\left(1+\cos2\gamma+\epsilon
\right)a^2\right]\sin\gamma\\
C&=&(1+\epsilon)(1+\epsilon a^2)H {\cal A}_{\rm CP}^{\rm mix}(B_d\to D^+D^-)
\cos\gamma,
\end{eqnarray}
with $a$ given in  (\ref{a-det}). Finally, $\sin\phi_d$ can be determined
as follows:
\begin{equation}
\sin\phi_d=\frac{AC-B\sqrt{A^2+B^2-C^2}}{A^2+B^2}.
\end{equation}
Here we have chosen the sign in front of the square root such that we obtain
a positive value of $\cos\phi_d$, in agreement with the $B$-factory data for 
the CP-violating effects in the $B_d\to J/\psi K^*$ and $B_d \to D^*D^*K_{\rm S}$ 
channels \cite{HFAG}. In Fig.~\ref{fig:Amix-s2b}, we show the resulting correlation
between the mixing-induced CP violation in $B^0_d\to D^+D^-$ and $\sin\phi_d$ 
for various values of $H$ and ${\cal A}_{\rm CP}^{\rm dir}(B_d\to D^+D^-)$,
which correspond to the situation shown in Fig.~\ref{fig:hadr-det}. These curves 
allow us straightforwardly to include the penguin effects in the determination of the 
$B^0_d$--$\bar B^0_d$ mixing phase form the CP-violating 
$B^0_d\to D^+D^-$ observables.

\boldmath
\subsection{Extraction of the $B^0_s$--$\bar B^0_s$ Mixing Phase}\label{ssec:phis-extr}
\unboldmath
Let us now turn to the CP-violating rate asymmetry of the $B^0_s\to D^+_sD^-_s$
decay, which is defined in analogy to (\ref{rate-asym}), and takes the form
\begin{eqnarray}
\lefteqn{{\cal A}_{\rm CP}(B_s(t)\to D^+_sD^-_s)}\nonumber\\
&&=\left[\frac{{\cal A}_{\rm CP}^{\rm dir}(B_s\to D^+_sD^-_s)\,\cos(\Delta M_s t)+
{\cal A}_{\rm CP}^{\rm mix}(B_s\to D^+_sD^-_s)\,\sin(\Delta M_s t)}{\cosh(\Delta\Gamma_st/2)-{\cal A}_{\rm \Delta\Gamma}(B_s\to D^+_sD^-_s)
\,\sinh(\Delta\Gamma_st/2)}\right],
\end{eqnarray}
where $\Delta\Gamma_s\equiv\Gamma_{\rm H}^{(s)}-\Gamma_{\rm L}^{(s)}$
is the difference of the decay widths $\Gamma_{\rm H}^{(s)}$ and 
$\Gamma_{\rm L}^{(s)}$ of the ``heavy" and ``light" mass eigenstates of the 
$B_s$ system, respectively. The mass difference $\Delta M_s$ was 
recently measured at the Tevatron \cite{D0-DMs,CDF-DMs}, with a value that 
is consistent with the SM expectation. On the other hand, this result still allows 
for large CP-violating NP contributions to $B^0_s$--$\bar B^0_s$ mixing (see, for 
instance, Refs.~\cite{BF-06,DMspapers}). In this case, the mixing phase $\phi_s$
would take a sizeable value, and would manifest itself also through significant 
mixing-induced CP violation in $B^0_s\to D^+_sD^-_s$ at LHCb. In the SM,
we have on the other hand a tiny phase of $\phi_s=-2\lambda^2\eta\approx-2^\circ$, 
where $\eta$ is another Wolfenstein parameter. 

Using the formalism discussed in Ref.~\cite{kitz}, (\ref{ABsDsDs-ampl}) yields
\begin{eqnarray}
{\cal A}_{\rm CP}^{\rm dir}(B_s\to D_s^+D_s^-)&=&
-\left[\frac{2\epsilon a'\sin\theta'\sin\gamma}{1+2\epsilon a'\cos\theta'\cos\gamma
+\epsilon^2a'^2}\right]\label{Adirs-expr}\\
{\cal A}_{\rm CP}^{\rm mix}(B_s\to D_s^+D_s^-)&=&
\left[\frac{\sin\phi_s+2\epsilon a'\cos\theta'\sin(\phi_s+\gamma)+
\epsilon^2a'^2\sin(\phi_s+2\gamma)}{1+2\epsilon a'\cos\theta'\cos\gamma+
\epsilon^2a'^2}\right]\label{Amixs-expr}\\
{\cal A}_{\Delta\Gamma}(B_s\to D_s^+D_s^-)&=&-\left[
\frac{\cos\phi_s+2\epsilon a'\cos\theta'\cos(\phi_s+\gamma)+
\epsilon^2a'^2\cos(\phi_s+2\gamma)}{1+2\epsilon a'\cos\theta'\cos\gamma+
\epsilon^2a'^2}\right],\label{ADGs-expr}
\end{eqnarray}
and (\ref{U-spin-rel1}) implies the following $U$-spin relation \cite{RF-BDD}: 
\begin{equation}
\frac{{\cal A}_{\rm CP}^{\rm dir}(B_s\to D_s^+D_s^-)}{{\cal A}_{\rm CP}^{\rm dir}
(B_d\to D^+D^-)}=-\epsilon H.
\end{equation}
Thanks to the suppression through the $\epsilon$ parameter in (\ref{Amixs-expr}), 
the penguin effects are significantly smaller than in the case of $B^0_d\to D^+D^-$. 
Nevertheless, since we are aiming at precision measurements, it is important to 
be able to control them. Since we may determine the penguin parameters $a$ and 
$\theta$ as we have discussed above, the $U$-spin relations in (\ref{U-spin-rel1})
allow us to include the penguin effects also in the determination
of $\phi_s$. It is instructive to perform an expansion in powers of $\epsilon a'$, 
which yields
\begin{equation}
\sin\phi_s={\cal A}_{\rm CP}^{\rm mix}(B_s\to D_s^+D_s^-)\mp
2\epsilon a'\cos\theta'\sin\gamma
\sqrt{1-{\cal A}_{\rm CP}^{\rm mix}(B_s\to D_s^+D_s^-)^2}+{\cal O}((\epsilon a')^2),
\end{equation}
where $\mp$ refers to $\mbox{sgn}(\cos\phi_s)=\pm1$. For strategies
to determine this sign, which is positive in the SM, see Refs.~\cite{RF-Bpipi-07,DFN}.
Using (\ref{ctheta-det}), the relevant hadronic parameter can straightforwardly
be fixed:
\begin{equation}
2\epsilon a'\cos\theta'\sin\gamma=(1-H)\tan\gamma+{\cal O}(a^2).
\end{equation}

Let us finally have a closer look at the observable
\begin{equation}
{\cal A}_{\Delta\Gamma}(B_s\to D_s^+D_s^-)=-\cos\phi_s
+2\epsilon a'\cos\theta'\sin\gamma\sin\phi_s+{\cal O}((\epsilon a')^2),
\end{equation}
which can be extracted from the following ``untagged" rate:
\begin{eqnarray}
\lefteqn{\langle \Gamma(B_s(t)\to D_s^+D_s^-)\rangle  \equiv  
\Gamma(B^0_s(t)\to D_s^+D_s^-)+ \Gamma(\bar B^0_s(t)\to D_s^+D_s^-)}\nonumber\\
& & \propto  e^{-\Gamma_st}\left[e^{+\Delta\Gamma_s t/2}R_{\rm L}(B_s\to D_s^+D_s^-)
+e^{-\Delta\Gamma_s t/2}R_{\rm H}(B_s\to D_s^+D_s^-)\right].\label{untagged}
\end{eqnarray}
Here $\Gamma_s$ denotes the average of the decay widths of the 
``heavy" and ``light" mass eigenstates of the $B_s$ system, and
\begin{eqnarray}
R_{\rm L}(B_s\to D_s^+D_s^-)&\equiv&
1-{\cal A}_{\Delta\Gamma}(B_s\to D_s^+D_s^-)=1+\cos\phi_s+{\cal O}(\epsilon a')
\stackrel{\rm SM}{\approx}2,\\
R_{\rm H}(B_s\to D_s^+D_s^-)&\equiv&1+{\cal A}_{\Delta\Gamma}(B_s\to D_s^+D_s^-)=
1-\cos\phi_s+{\cal O}(\epsilon a')\stackrel{\rm SM}{\approx}0.
\end{eqnarray}
As far as a practical measurement of (\ref{untagged}) is concerned, most of
the data come from short times with $\Delta\Gamma_st\ll1$. We may hence
expand in this parameter, which yields
\begin{equation}
\langle \Gamma(B_s(t)\to D_s^+D_s^-)\rangle\propto
e^{-\Gamma_st}\left[1-{\cal A}_{\Delta\Gamma}(B_s\to D_s^+D_s^-)
\left(\frac{\Delta\Gamma_st}{2}\right)+{\cal O}((\Delta\Gamma_st)^2)\right].
\end{equation}
Moreover, if the two-exponential form of (\ref{untagged}) is fitted to a single 
exponential, the corresponding decay width satisfies the following relation \cite{DFN}:
\begin{equation}\label{Gamma-fit}
\Gamma_{D_s^+D_s^-}=\Gamma_s+{\cal A}_{\Delta\Gamma}(B_s\to D_s^+D_s^-)
\frac{\Delta\Gamma_s}{2}+{\cal O}((\Delta\Gamma_s)^2/\Gamma_s).
\end{equation}
Using flavour-specific $B_s$ decays, a similar analysis allows the extraction 
of $\Gamma_s$ up to corrections of ${\cal O}((\Delta\Gamma_s/\Gamma_s)^2)$
\cite{DFN}. In the presence of NP, $\Delta\Gamma_s$ is modified as follows 
\cite{grossman}:
\begin{equation}
\Delta\Gamma_s=\Delta\Gamma_s^{\rm SM}\cos\phi_s,
\end{equation}
where $\Delta\Gamma_s^{\rm SM}/\Gamma_s$ is negative for the definition 
given above, and calculated at the 15\% level~\cite{LN}. Consequently, 
(\ref{Gamma-fit}) actually probes
\begin{equation}
\Gamma_{D_s^+D_s^-}-\Gamma_s=\left[\cos^2\phi_s
-\epsilon a'\cos\theta'\sin(2\phi_s)\right]\frac{|\Delta\Gamma_s^{\rm SM}|}{2} + \ldots,
\end{equation}
thereby complementing other determinations of the width difference of the
$B_s$ system, such as from the $U$-spin-related $B_s\to K^+K^-$, $B_d\to\pi^+\pi^-$ 
decays \cite{RF-Bpipi-07}.

\boldmath
\section{Interplay with Other Probes of CP Violation}\label{sec:inter}
\unboldmath
As we have seen in the previous section, the $U$-spin-related $B_q\to D_q^+D_q^-$ 
decays offer an interesting tool for the extraction of the $B^0_q$--$\bar B_q$ 
mixing phases ($q\in\{d,s\}$). Since the ``golden" decay $B^0_d\to J/\psi K_{\rm S}$ 
and similar channels allow already a very impressive determination of $\phi_d$,
as can be seen in (\ref{phid-exp}), this may not look as too exciting. However, this 
is actually not the case. In fact, the current value of (\ref{phid-exp}) is on the lower 
side, and the interplay with the UT side $R_b\propto |V_{ub}/V_{cb}|$ leads to some 
tension in the CKM fits \cite{HFAG,UTfit,CKMfitter}, which receives increasing attention
in the $B$-physics community. If this effect is attributed to NP, the standard
interpretation is through CP-violating contributions to $B^0_d$--$\bar B^0_d$
mixing, with a NP phase $\phi_d^{\rm NP}\sim -10^\circ$ \cite{NP-Bd,BF-06}. 

However, the NP effects could also enter through the $B^0_d\to J/\psi K_{\rm S}$
amplitude, where EW penguin topologies, which have a sizeable
impact on this decay \cite{EWP-rev}, offer a particularly interesting scenario. 
The $B$-factory data for $B\to\pi\pi, \pi K$ modes may actually indicate a modified 
EW penguin sector with a large CP-violating NP phase through the results for 
mixing-induced CP violation in $B^0_d\to\pi^0 K_{\rm S}$ \cite{FRS,BFRS}, 
thereby complementing the pattern of such CP asymmetries observed in other 
$b\to s$ penguin modes, where the $B^0_d\to\phi K_{\rm S}$ channel is an 
outstanding example \cite{HFAG}. The sign of the corresponding CP-violating 
NP phase would actually shift ${\cal A}_{\rm CP}^{\rm mix}(B_d\to J/\psi K_{\rm S})$ 
in the right direction \cite{FRS,EWP-rev}. The interesting feature of the 
$B_{d(s)}\to D_{d(s)}^+D_{d(s)}^-$ decays is that they are essentially unaffected 
by such a NP scenario as EW penguins contribute only in colour-suppressed 
form and play a minor r\^ole. Consequently, a difference between the values 
of $\phi_d$ extracted from $B_d\to  J/\psi K_{\rm S}$ and the 
$B_{d(s)}\to D_{d(s)}^+D_{d(s)}^-$ system could reveal such effects. 

A similar comment applies to the determination of the $B^0_s$--$\bar B^0_s$
mixing phase, where the ``golden" strategy uses mixing-induced CP violation 
in the time-dependent angular distribution of the 
$B_s\to J/\psi[\to\ell^+\ell^-]\phi[\to K^+K^-]$ decay products 
\cite{DDF,DFN}; penguin effects can be controlled with the help of
$B_d\to J/\psi\rho^0$ \cite{ang}. This determination of
$\phi_s$ could also be affected by CP-violating NP contributions entering through 
EW penguin topologies. On the other hand, the extraction discussed in 
Subsection~\ref{ssec:phis-extr} is essentially unaffected, so that a difference
between the two results could again signal such a kind of physics beyond the SM. 
Moreover, also a simultaneous analysis of the $U$-spin-related 
$B_{s(d)}\to J/\psi K_{\rm S}$ decays should be performed \cite{RF-BDD}.  In 
analogy to the discussion given above, the (small) penguin effects in the determination
of $\phi_d$ from $B_d\to  J/\psi K_{\rm S}$ can then be controlled, and $\phi_s$
could be extracted  from the $b\to d$ channel $B_{s}\to J/\psi K_{\rm S}$, again with 
a sensitivity to a modified CP-violating EW penguin sector.

As was noted in Ref.~\cite{habil}, the analysis of the $B_{d(s)}\to D^+_{d(s)}D^-_{d(s)}$
decays can also straightforwardly be applied to the $B_{d(s)}\to K^0 \bar K^0$ system.
Following these lines, the penguin effects in the determination of
$\sin\phi_s$ from the $b\to s$ penguin decay $B^0_s\to K^0\bar K^0$
can be included through its $B^0_d\to K^0\bar K^0$ partner \cite{CPS};\footnote{Here
$B^0_s\to K^0\bar K^0$ and $B^0_d\to K^0\bar K^0$ take the r\^oles of 
$B^0_s\to D^+_sD^-_s$ and $B^0_d\to D^+D^-$, respectively.}
this is also the case for the corresponding $B_{d(s)}\to K^{*0} \bar K^{*0}$ decays 
\cite{ang,matias}. Again in these transitions, EW penguin have a very small impact.
Should the interesting pattern in the mixing-induced CP asymmetries of 
$B^0_d\to \pi^0 K_{\rm S}$, $B^0_d\to \phi K_{\rm S}$ and similar modes originate 
from a modified EW penguin sector, we would again not see it in the 
$B_{d(s)}\to K^{(*)0} \bar K^{(*)0}$ system.

\boldmath
\section{Conclusions}\label{sec:concl}
\unboldmath
The CP violation in $B^0_d\to D^+D^-$ offers another interesting probe 
for the exploration of the Kobayashi--Maskawa mechanism of CP violation. In these
studies, the penguin effects have to be controlled, which can be done with the
help of the $U$-spin-related $B^0_s\to D_s^+D_s^-$ channel. Motivated by the
recent data from the $B$ factories and the Tevatron, as well as the quickly 
approaching start of the LHC, we have investigated 
the allowed region in the space of the mixing-induced and direct CP violation 
of the $B^0_d\to D^+D^-$ decay, with useful results to monitor the future 
improvement of the experimental picture, and have performed theoretical 
estimates of the relevant hadronic parameters and observables. 

We then discussed the extraction of CP-violating phases, where we may either use 
$\phi_d$ as an input to determine $\gamma$, or use $\gamma$ to extract $\phi_d$. 
Concerning the former option, the current data point towards an unstable situation for 
the extraction of $\gamma$, while the strong phase $\theta$ could be well determined. 
It appears therefore more interesting to extract the $B^0_d$--$\bar B^0_d$ mixing phase 
from the CP asymmetries of $B^0_d\to D^+D^-$, also since precision measurements 
of $\gamma$ will be available from the LHCb experiment through other strategies. 
We have provided the formalism to include the penguin effects, and have 
illustrated its practical implementation. In the case of the CP asymmetries of the 
$B^0_s\to D_s^+D_s^-$ decay, the penguin effects are doubly Cabibbo-suppressed 
and play therefore a significantly less pronounced r\^ole. However, they can also be 
taken into account with the help of the $B^0_d\to D^+D^-$ decay, allowing then a 
precision measurement of the $B^0_s$--$\bar B^0_s$ mixing phase from the 
mixing-induced CP violation in $B^0_s\to D_s^+D_s^-$.

An interesting feature of these determinations is the fact that they are insensitive
to CP-violating NP contributions entering through the EW penguin sector. In this
respect, they are complementary to the well-known standard strategies. The 
determinations of the $B^0_s$--$\bar B^0_s$ mixing phase through the 
$B_{s(d)}\to D_{s(d)}^+D_{s(d)}^-$ system on the one hand and 
$B_s\to J/\psi \phi$, $B_d\to J/\psi\rho^0$ on the other hand are particularly 
promising, and the studies of LHCb in this direction should be further pursued
to fully exploit the physics potential of these decays.

\end{document}